\documentclass[%
 reprint,
superscriptaddress,
twocolumn,
 amsmath,amssymb,
 aps,
 pre,
]{revtex4-2}

\usepackage{amssymb,amsmath,bm}
\usepackage{enumerate}
\usepackage[utf8]{inputenc}
\usepackage{cancel}
\usepackage{ulem}
\usepackage{color}
\usepackage[colorlinks,linkcolor=blue,urlcolor=blue,citecolor=blue]{hyperref}
\usepackage{float}
\usepackage{xr}
\usepackage[toc]{appendix}
\usepackage{graphicx,epstopdf}
\usepackage{tikz}

\usepackage{mathabx}

\def\be{\begin{equation}}
\def\ee{\end{equation}}
\def\bea{\begin{eqnarray}}
\def\eea{\end{eqnarray}}

\begin{document}
\title{Effect of reactive targets on diffusive transport with intermittent restarts}
\author{Samali Ghosh}
\email{samalighosh816@gmail.com}
\affiliation{Physics and Applied Mathematics Unit, Indian Statistical Institute, 203 B.T. Road, Kolkata, 700108, India}
\author{Suvam Pal}
\email{suvamjoy256@gmail.com}
\affiliation{Physics and Applied Mathematics Unit, Indian Statistical Institute, 203 B.T. Road, Kolkata, 700108, India}
\author{Dibakar Ghosh}
\email{dibakar@isical.ac.in}
\affiliation{Physics and Applied Mathematics Unit, Indian Statistical Institute, 203 B.T. Road, Kolkata, 700108, India}
\author{P. S. Pal}
\email{pspal@kias.re.kr}
\affiliation{School of Computational Sciences, Korea Institute for Advanced Study, Seoul 02455, Korea}

\begin{abstract} 
    We study search processes in a complex environment using the strategy of stochastic resetting. A common occurrence in these environments is targets with finite reactivity. Stochastic resetting has emerged as a powerful mechanism for optimizing search processes by curtailing long, unproductive excursions inherent to diffusive dynamics. Most theoretical studies, however, assume  perfectly absorbing targets -- an idealization that overlooks the finite reactivity commonly encountered in realistic chemical and biological systems. In this work, we investigate the interplay between stochastic resetting and finite-target reactivity in reaction - diffusion processes. Considering a one-dimensional system with multiple reactive targets, we demonstrated that the target reactivity modifies the optimization landscape. We uncover distinct regimes in which resetting enhances transport efficiency towards specific targets based on their chemical kinetics. As a consequence, the optimal resetting rate becomes intrinsically reactivity dependent. Our results identify target reactivity  as a crucial control parameter governing stochastic transport and provide insights into reaction-diffusion processes in complex media.

    
\end{abstract}

\maketitle

\section{Introduction}

Search processes in complex environments underlie a wide range of phenomena in physics, chemistry, and biology which includes chemical reactions, intracellular transport, and molecular recognition. In such systems an agent - a diffusing particle or a chemical reactant - navigate through the media to locate a specific target where a reaction or transformation can occur. The efficiency of such processes is not only determined by the stochastic dynamics that the agent goes through but also on the properties of the target.
A central question concerns the time required for a randomly moving particle to reach a specified set of locations for the first time. This question defines the class of first-passage processes, in which the focus shifts from the detailed trajectory of motion to the statistics of the earliest arrival at one or more targets. In many realistic situations, the environment contains multiple competing targets, such as reactive sites, traps, or exits, each of which can terminate the dynamics upon encounter. The presence of multiple targets introduces a competition between distinct first-passage channels, leading to nontrivial splitting probabilities, conditional first-passage times (FPTs), and correlations between target geometry and transport dynamics\cite{rostovtseva2008tubulin,hoogerheide2017mechanism,dagdug2024diffusion,jain2023fick,pal2024channel}. As a result, first-passage problems with multiple targets provide a powerful framework for quantifying search efficiency, reaction selectivity, and pathway competition in stochastic transport processes.
%

%
A common assumption in theoretical studies is that the targets are \textbf{ perfectly absorbing}, such that the process/dynamics terminates immediately upon first encounter with a target. In realistic settings, however, successful completion may require additional microscopic conditions, such as proper orientation, conformational alignment, or the crossing of an energy barrier. Consequently, not every encounter results in successful absorption, and such targets are more appropriately described as \textbf{ partially absorbing}~\cite{erban2007reactive,pal2019motion,bressloff2025macroscopic}. Partially absorbing targets are commonly modeled using radiation (or Robin) boundary condition \cite{erban2007reactive}, under which only a fraction of incoming diffusive flux is absorbed at the target, while the remainder is transferred to the bulk. The associated intrinsic reaction rate characterizes the reactivity of the target. The relevance of reactive targets extends across a variety of transport and reaction processes encountered in surface and colloid science and materials research~\cite{lomholt2007subdiffusion}. Examples range from transport through porous interfaces~\cite{stapf1995proton}, plasma etching~\cite{vrentas1989boundary} to electric transport in electrolytic cells~\cite{sapoval1994general} and diffusion-controlled reaction at catalytic surfaces~\cite{sapoval2001catalytic,andrade2001analytical,andrade2003transition}.
%
%
%

%
In the context of first passage problems~\cite{redner2001guide}, stochastic resetting~\cite{evans2011diffusion,evans2020stochastic} has emerged over the past decade as a powerful tool to control and optimize stochastic processes.  In this framework, the dynamics is intermittently interrupted, and state of the agent (e.g., diffusive particle) is reinitialized, thereby suppressing long unproductive excursions. 
Researchers have extensively studied various aspects of stochastic resetting, including properties of its nonequilibrium steady state~\cite{evans2011diffusion,eule2016non,pal2016diffusion,evans2020stochastic,tal2020experimental,gupta2022stochastic} and methods to further accelerate search processes~\cite{pal2017first,ray2021mitigating,mendez2022nonstandard,sar2023resetting,pal2025universal,pal2025optimal,biswas2025target,pal2026resetting}.
Beyond its applications in physics, stochastic resetting also plays a crucial role in accelerating various biological processes. Examples include kinetic proofreading ~\cite{bar2002protein,murugan2012speed,pal2019landau}, protein-folding process ~\cite{bhaskaran2007kinetic,hyeon2013generalized,chakrabarti2017molecular}, and chemical reaction process ~\cite{reuveni2014role,biswas2023rate}. See \cite{evans2020stochastic,pal2022inspection,pal2024random} for an extensive review on the topic. 

\begin{figure}
    \centering
    \includegraphics[width=\linewidth]{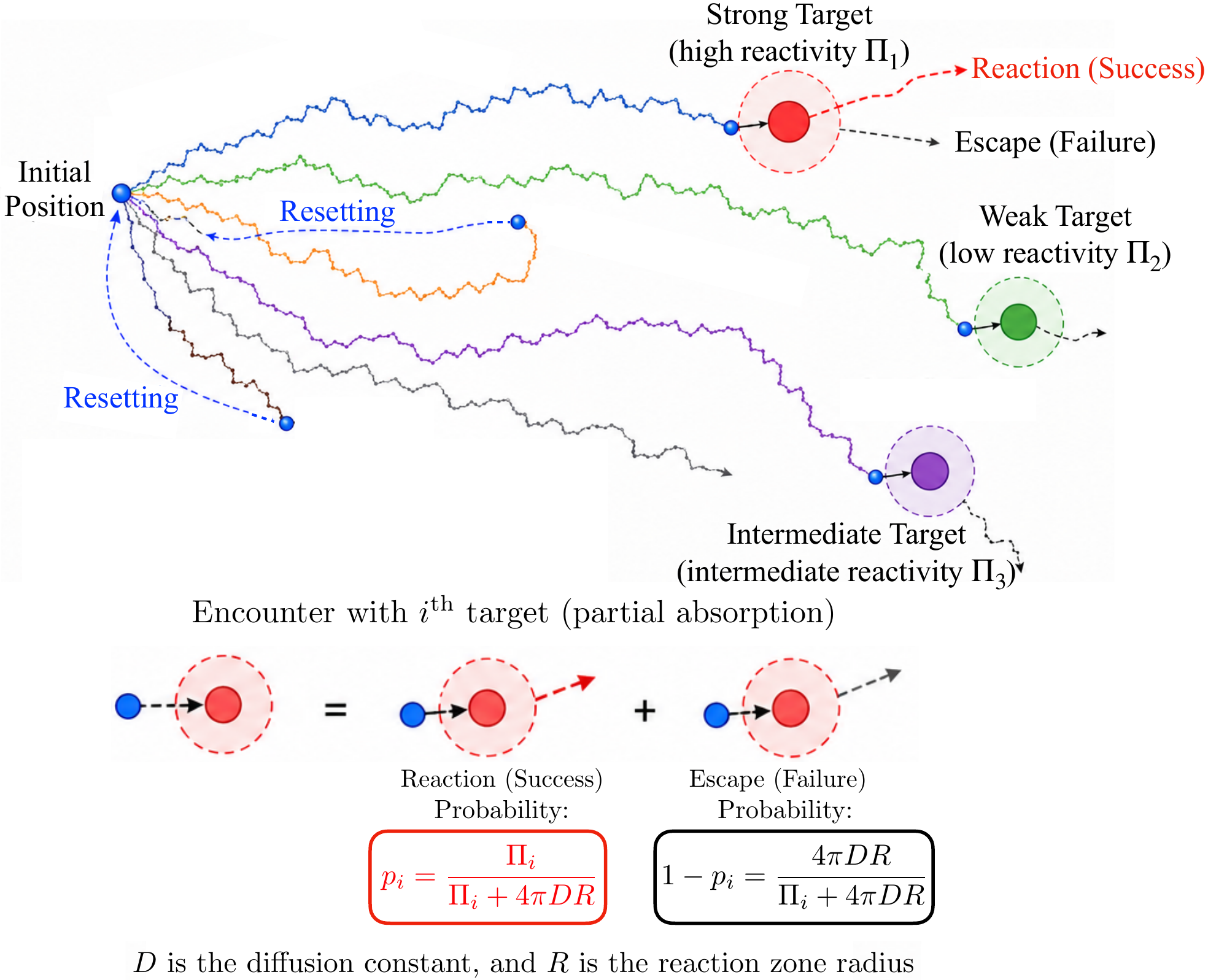}
    \caption{Schematics of a resetting-induced first-passage process in the presence of multiple reactive targets. A substrate molecule searches for enzyme molecules acting as partially absorbing targets with different reactivities $\Pi_i$. Upon encounter with the target, the molecule can either react (absorption) or escape without absorption, reflecting the finite reactivity of the targets. Stochastic resetting randomly relocates the molecule, generating a competition between diffusion, resetting, and target reactivity that governs the search efficiency. The lower panel illustrates the reaction and escape probabilities associated with a partially absorbing target.}
    \label{schematic_diagram}
\end{figure}

A typical example is provided by a diffusion-limited enzyme-substrate system in a solution (Fig.~\ref{schematic_diagram}). The substrate molecules diffuse in the solution to locate the enzyme molecules, which act as reactive targets with their intrinsic reactivity determined by catalytic efficiency or conformational state. Upon reaching the vicinity of the enzyme, successful conversion is not guaranteed, and the substrate molecule might escape back to the bulk. As a result, multiple encounters might be required before a successful absorption to occur. The targets are therefore practically partially absorbing. In such systems, stochastic resetting can arise intrinsically as an effective description of intermittent processes that interrupt the search processes, such as relocation due to environmental fluctuations, binding-unbinding with other structures, etc. These processes lead to a loss of correlation with the previous trajectory, effectively reinitializing the search from a new configuration, thereby generating a renewal structure analogous to stochastic resetting.

In this paper, we mimic the aforementioned scenario through a reaction-diffusion process under intermittent resetting in one dimension bounded by two reactive targets and investigate its first-passage properties. The combined effect of reactive confinement and stochastic resetting reveals a nontrivial coupling between bulk intermittency and interfacial kinetics, enriching the phenomenology of first-passage processes beyond ideal absorbing scenarios. The manuscript is structured as follows. In Sec.~\ref{sec2}, we introduce the mathematical framework for our system. Following this, in Sec.~\ref{sec3} we extend our discussion in the presence of resetting. Starting with the survival probability, we evaluate the mean first-passage time (MFPT) for both unconditional and conditional outcomes.

\section{System in absence of resetting} \label{sec2}
Consider a Brownian particle diffusing inside a one-dimensional box of length $l$, bounded by two reactive walls with reactivities $\Pi_0$ and $\Pi_l$ at the left and right boundaries, respectively. Starting from an initial position $x_0 \in (0, l)$, the particle undergoes stochastic motion and may eventually reach one of the boundaries. Upon reaching a particular boundary, the probability of absorption is determined by the reactivity of the corresponding boundary, thereby terminating the particle's trajectory; with the complementary probability, it is reflected into the domain, allowing the motion to continue. This probabilistic interplay between absorption and reflection at the boundaries plays a crucial role in determining the first-passage time (FPT) statistics of the system.

In particular, the evolution of the probability density function $p_0(x,t)$ is governed by the Fokker-Planck equation
\begin{equation}\label{fp}
    \dfrac{\partial}{\partial t}p_0(x,t)=D\dfrac{\partial^2}{\partial x^2}p_0(x,t),
\end{equation}
where the diffusion constant is denoted by $D$ and the initial condition $p_0(x,0)=\delta(x-x_0)$.  Eq.~\eqref{fp} can be solved by imposing  the Robin boundary conditions 
\begin{align}
    &\left.D\partial_x p_0(x,t)\right|_{x\rightarrow 0}=\Pi_0~p_0(0,t),\label{left-bc}\\
    &\left.D\partial_x p_0(x,t)\right|_{x\rightarrow l}=-\Pi_l~p_0(l,t).\label{right-bc}
\end{align} 
Furthermore, in the limit $\Pi_{0(l)}\to \infty$, the boundary becomes perfectly absorbing, resulting in a vanishing probability density at that boundary. Conversely, $\Pi_{0(l)}\to 0$ corresponds to a zero-flux condition and hence to a perfectly reflecting boundary with no loss of probability.
Notably, Eqs.~\eqref{left-bc} and \eqref{right-bc} characterize the partially absorbing and partially reflecting nature of the reactive boundaries. Moreover, taking the Laplace transformation of Eq.~\eqref{fp}, one obtains
\begin{equation}\label{fp-lap}
    s\widetilde{p}_0(x,s)-\delta(x-x_0)=D\dfrac{\partial^2}{\partial x^2}\widetilde{p}_0(x,s).
\end{equation}
Here $\widetilde{z}(s)$ is denoted as the Laplace-transformed quantity, \textit{i.e.,} $\widetilde{z}(s)=\mathcal{L}[z(t)]=\int_0^\infty e^{-st}z(t) dt$. Imposing the boundary conditions from Eqs.~\eqref{left-bc} and \eqref{right-bc}, the solution of Eq.~\eqref{fp-lap} in the Laplace domain is given by
\begin{align}\label{fp-sol}
    \widetilde{p}_0(x,s)&=\dfrac{1}{\alpha(s)D\mathcal{N}(s)}\times \begin{cases}
        \mathcal{W}(x,x_0,s), & 0 \leq x \leq x_0, \\
        \mathcal{W}(x,x_0,s), & x_0 \leq x \leq l,
    \end{cases}
\end{align}
with the following functional expressions 
\begin{align}
    \mathcal{W}&(x_0,x,s)\nonumber\\
    &=\Big(\alpha(s)D\cosh[(l-x_0)\alpha(s)]+\Pi_l\sinh[(l-x_0)\alpha(s)]\Big)\nonumber\\
    &~~~~\times \Big(\alpha(s)D\cosh[x\alpha(s)]+\Pi_0\sinh[x\alpha(s)]\Big), \nonumber\\
    \mathcal{N}&(s)=\Big(\alpha(s)D(\Pi_0+\Pi_l)\cosh[l\alpha(s)]\nonumber\\
    &~~~~~~~~~~~~~~~~~~~~~~+(\Pi_0\Pi_l+\alpha^2(s)D^2)\sinh[l\alpha(s)]\Big),\nonumber
\end{align}
where $\alpha(s)=\sqrt{s/D}$. The probability fluxes through the left (-) and right (+) boundaries are defined as $J^\mp_0(x_0,t)=\pm D \left.\partial_xp_0(x,t)\right|_{x\to 0(l)}$. Setting the initial position to $x_0=ul$ and introducing the reactivity ratio $n=\Pi_l/\Pi_0$, the corresponding fluxes in the Laplace domain are obtained as
\begin{align}
    &\widetilde{J}^-_0(u,s)=\dfrac{\mathcal{G}(\Pi_0,\Pi_l,u,s)}{\mathcal{N}(s)},\label{j0-minus}\\
    &\widetilde{J}^+_0(u,s)=\dfrac{\mathcal{G}(\Pi_l,\Pi_0,1-u,s)}{\mathcal{N}(s)},\label{j0-plus}
\end{align}
with $\mathcal{G}(k_1,k_2,u,s)=k_1\Big(\alpha(s)D\cosh[(1-u)l\alpha(s)]+k_2\sinh[(1-u)l\alpha(s)]\Big)$. 
The escape (or splitting) probabilities are related to the probability fluxes through the respective targets via $\epsilon^\mp_0(u)=\int_0^\infty J^\mp_0(u,t)\,dt$ which, in the Laplace domain, becomes $\epsilon_0^\mp(x_0)=\widetilde{J}^\mp_0(x_0,s\rightarrow 0)$. Using this relation, we obtain
\begin{align}
    \epsilon_0^-(u)&=\mathcal{F}(\Pi_0,\Pi_l,u),\label{eps0-minus}\\
    \epsilon_0^+(u)&=\mathcal{F}(\Pi_l,\Pi_0,1-u),\label{eps0-plus}
\end{align}
where $\mathcal{F}(k_1,k_2,u)=k_1(D+k_2l(1-u))/(D(k_1+k_2)+k_1k_2l)$. Importantly, in the limit $\Pi_{0(l)}\to \infty$,  the splitting probabilities reduce to those for purely absorbing boundaries, i.e., $\epsilon_0^-(u)=1-u$ and $\epsilon_0^+(u)=u$. Moreover, the conditional FPT densities corresponding to the absorption at the left and right targets are related to the associated probability fluxes as $f_{T^\mp_0}(u,t)=J^\mp_0(u,t)/\epsilon_0^\mp(u)$. Consequently, the conditional FPT densities in the Laplace domain are given by
\begin{align}
    &\widetilde{T}^-_0(u,s)=\dfrac{\widetilde{J}^-_0(u,s)}{\epsilon_0^-(u)}=\dfrac{1}{\mathcal{N}(s)}\dfrac{\mathcal{G}(\Pi_0,\Pi_l,u,s)}{\mathcal{F}(\Pi_0,\Pi_l,u)},\label{fpt0-minus}\\
    &\widetilde{T}^+_0(u,s)=\dfrac{\widetilde{J}^+_0(u,s)}{\epsilon_0^+(u)}=\dfrac{1}{\mathcal{N}(s)}\dfrac{\mathcal{G}(\Pi_l,\Pi_0,1-u,s)}{\mathcal{F}(\Pi_l,\Pi_0,1-u)}.\label{fpt0-plus}
\end{align}
The corresponding mean values of conditional FPT are
\begin{align}
    &\langle T^-_0(u)\rangle = \Gamma (u,\Pi_0,\Pi_l),\label{mfpt0-minus}\\
    &\langle T^+_0(u)\rangle = \Gamma (1-u,\Pi_l,\Pi_0),\label{mfpt0-plus}
\end{align}
respectively. Here, $\Gamma(u,\Pi_0,\Pi_l)=l (6 D^3-3 D^2 l \left(\Pi_0 (u-2) u+\Pi_l \left(u^2-2\right)\right)+D \Pi_l l^2 (\Pi_0 u ((u-6) u+6)+\Pi_l ((u-3) u^2+2))+\Pi_0 \Pi_l^2 l^3 (u-2) (u-1) u)/(6 D (D (\Pi_0+\Pi_l)+\Pi_0 \Pi_l l) (D+\Pi_l l (1-u)))$. Note that, when the targets become perfectly absorbing, represented by the limits $\Pi_0\to \infty$, and $\Pi_l\to \infty$, Eqs.~\eqref{mfpt0-minus} and \eqref{mfpt0-plus} simplify to $\langle T^-_0(u)\rangle=(l^2/6D)u(2-u)$ and $\langle T^+_0(u)\rangle = (l^2/6D) (1-u^2)$~\cite{redner2001guide}. 
In addition, the unconditional FPT density can be expressed as  the weighted sum of the conditional FPT densities, yielding
\begin{align}
    \widetilde{T}_0(u,s)&=\epsilon_0^-(u)\widetilde{T}^-_0(u,s)+\epsilon_0^+(u)\widetilde{T}^+_0(u,s)\nonumber\\
    &=\dfrac{1}{\mathcal{N}(s)}\left[\mathcal{G}(\Pi_0,\Pi_l,u,s)+\mathcal{G}(\Pi_l,\Pi_0,1-u,s)\right].\label{fpt0}
\end{align}
The mean unconditional FPT is then given by
    \begin{align}\label{mfpt0}
        \langle T_0(u)\rangle &= \dfrac{l}{2D[D(\Pi_0+\Pi_l)+\Pi_0\Pi_ll]}\times\nonumber\\
        &~~~~~~~~~~\bigg[2D^2+\Pi_0\Pi_lu(1-u)l^2\nonumber\\
        &\hspace{1.2cm}+D\{\Pi_l+2\Pi_0u-(\Pi_0+\Pi_l)u^2\}l\bigg].
    \end{align}
Importantly, in the limit $\Pi_0\to \infty$ and $\Pi_l\to \infty$, Eq.~\eqref{mfpt0} reduces to $\langle T_0(u)\rangle=(l^2/2D)u(1-u)$~\cite{redner2001guide}.

\section{System dynamics in presence of resetting} \label{sec3}
Having characterized the underlying dynamics without resetting, we now turn to the study of first passage dynamics in presence of stochastic resetting, where the dynamics is intermittently interrupted and reinitiated from the initial configuration. Such resetting suppresses excessively long excursions, thereby promoting a more efficient exploration of the phase space. Stochastic resetting also generates a renewal structure in the dynamics that enables the key observables--such as splitting probabilities, probability fluxes, and FPT densities-- to be expressed in terms of the corresponding observables of the underlying dynamics without resetting.
\subsection{Unconditional first-passage process in presence of stochastic resetting}
\begin{figure}
    \centering
    \includegraphics[width=1.03\linewidth]{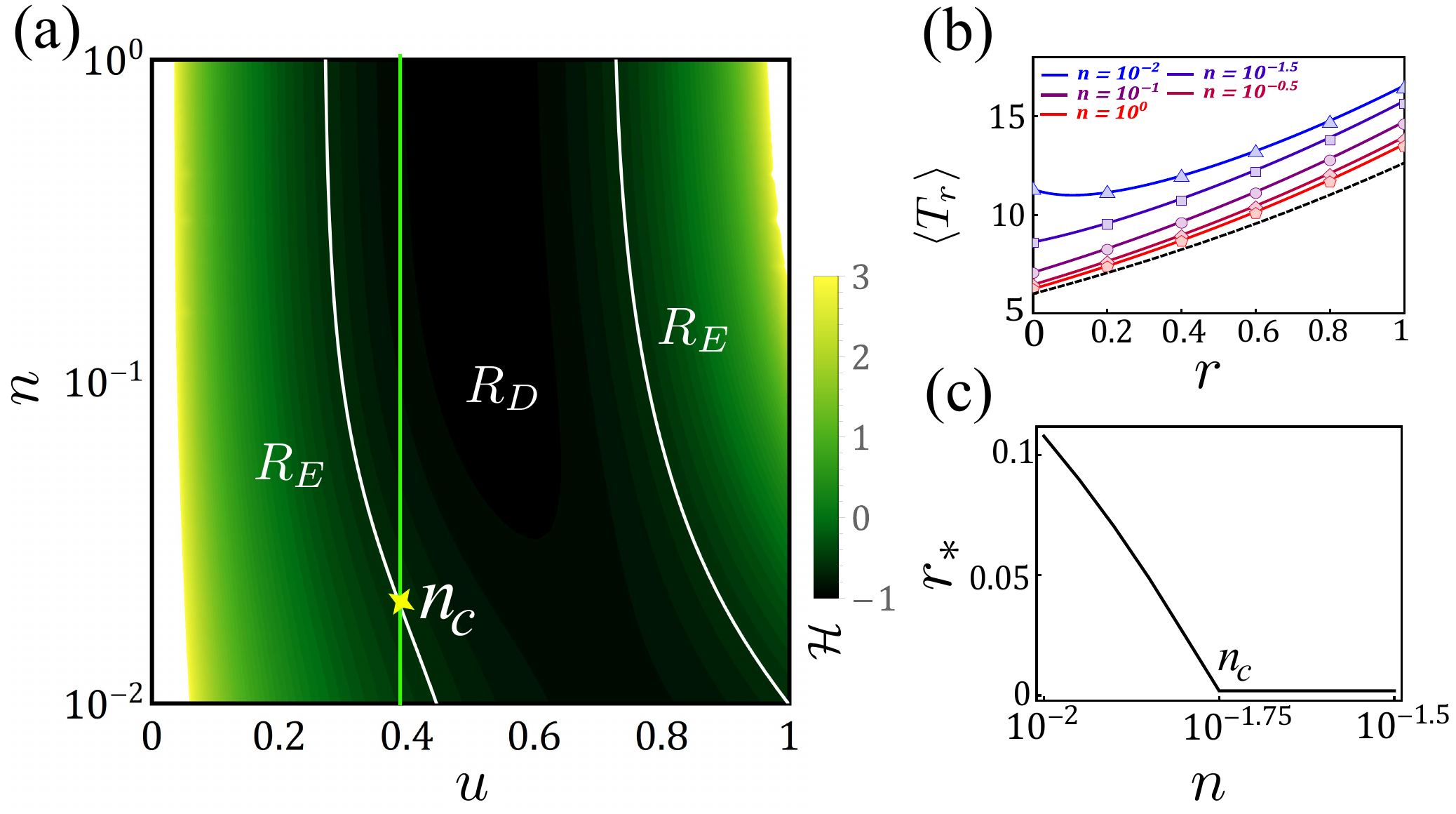}
    \caption{
    (a) Heat map of $\mathcal{H}(n,\Pi_0,u,l)$ in $(n-u)$ space. The white contours correspond to $\mathcal{H}(n,\Pi_0,u,l)=0$, or equivalently $CV_0(u)=1$. These contours separate the parameter space into the resetting-enhanced regime $R_{\rm E}$ and the detrimental-resetting region $R_{\rm D}$.
    (b) Unconditional MFPT as a function of the resetting rate for different values of $n$, where the particle is initially at $u= 0.4$ [also marked in (a)]. The black dashed line represents the unconditional MFPT in presence of the purely absorbing boundaries. 
    (c) Optimal resetting rate $r_*$ as a function of reactivity ratio $n$. The critical reactivity ratio $n_c$ as obtained from this plot is also marked in (a).
    Parameters: $l=5,~D=0.5$ and $\Pi_0=10$.
    }
    \label{fig2}
\end{figure}

We begin by introducing the survival probability $Q_{r}(u,t)$ defined as the probability that the particle starting from $u$ has not been absorbed at either boundary up to time $t$ in the presence of intermittent restarts at random epochs. Here, survival refers to the absence of absorption rather than the absence of encounters with the targets, since the encounter with a partially reactive boundary does not necessarily terminate the process. The resetting events occur independently with exponentially distributed waiting times of rate $r$. The intervals between successive resetting events are assumed to be independent and drawn from the \textit{exponential} distribution with rate $r$.  In terms of the underlying survival probability $Q_0(u,t)$, the last renewal equation takes the form
\begin{align} \label{surv-renew}
    Q_r(u,t)=&e^{-rt}Q_0(u,t)\nonumber\\
    &+r\int_0^t e^{-r\tau}~Q_r(u,t-\tau)~Q_0(u,\tau)~d\tau,
\end{align}
where the first term, composed of two factors, represents the probability that no resetting event occurs up to time $t$. Specifically, $e^{-rt}$ is the probability that the dynamics evolve without resetting over the interval $[0,t]$, while $Q_0(u,t)$ denotes the survival probability in the absence of resetting.
Additionally, $Q_0(u,t)=1-\int_0^tf_{T_0}(u,\tau)~d\tau$, with $f_{T_0}(u,t)$ being the corresponding unconditional FPT density. 
The second term captures trajectories with one or more resetting events before time $t$. In particular, the last resetting occurs within the interval  $[t-\tau,t-\tau+d\tau]$, with probability $re^{-r\tau}d\tau$,  while the particle survives up to that event with  probability $Q_r(u,t-\tau)$. Subsequently, over the remaining time interval $\tau$, the dynamics evolve without resetting and survive with probability $Q_0(u,\tau)$. Integration over all possible $\tau$ yields the total contribution from such renewal events. Taking the Laplace transform of Eq.~\eqref{surv-renew}, we obtain
\begin{equation}\label{surv-renew-lap}
    \widetilde{Q}_r(u,s)=\dfrac{\widetilde{Q}_0(u,r+s)}{1-r\widetilde{Q}_0(u,r+s)},
\end{equation}
where $\widetilde{Q}_r(u,s)=\int_0^\infty~e^{-st}Q_r(u,t)~dt$. Furthermore, using the relation in the Laplace domain between unconditional FPT density and the survival probability, i.e., $\widetilde{T}_{r(0)}(u,s) = 1-s\widetilde{Q}_{r(0)}(u,s)$, the resetting-induced unconditional FPT density can be expressed as
\begin{equation}\label{uncond-fpt-res-lap-1}
    \widetilde{T}_r(u,s)=\dfrac{(r+s)\widetilde{T}_0(u,r+s)}{s+r\widetilde{T}_0(u,r+s)}.
\end{equation}
The corresponding mean and second moment are given by
\begin{align}
    \langle T_r(u)\rangle &= \frac{(1-\widetilde{T}_0(u,r))}{r\widetilde{T}_0(u,r)},\nonumber\\
    \langle T_r(u)^2\rangle&=\frac{2(1-\widetilde{T}_0(u,r)+r\partial_r \widetilde{T}_0(u,r))}{r^2\widetilde{T}^2_0(u,r)}.
\end{align}
Using Eq.~\eqref{fpt0}, unconditional MFPT is obtained as
\begin{widetext}
    \begin{equation}\label{mfpt-uncond-res-main}
        \langle T_r(u)\rangle = \dfrac{1}{r}\left[\dfrac{D(\Pi_0+\Pi_l)\alpha(r)\cosh[l\alpha(r)]+(\Pi_0\Pi_l+D^2\alpha(r)^2)\sinh[l\alpha(r)]}{D\alpha(r)(\Pi_0\cosh[(1-u)l\alpha(r)]+\Pi_l\cosh[ul\alpha(r)])+\Pi_0\Pi_l\sinh[ul\alpha(r)]+\sinh[(1-u)l\alpha(r)]}-1\right].
    \end{equation}
\end{widetext}
Notably, with purely absorbing boundaries, \textit{i.e.,} $\Pi_0,\Pi_l \rightarrow\infty$, from Eq.~\eqref{mfpt-uncond-res-main}, one can recover \cite{pal2019first}, 
\begin{equation}\label{uncond-mfptr-abs}
    \langle T_r(u)\rangle = \dfrac{1}{r}\left[\dfrac{\cosh[\alpha(r)l/2]}{\cosh[\alpha(r)l(1-2u)/2]}-1\right].
\end{equation}
Intermittent restarts curtail long and unproductive excursions, thereby accelerating the completion of the search process. Conversely, excessively frequent resetting localizes the trajectories in the vicinity of the initial location $u$, reducing the likelihood of reaching the boundary. This competition can be characterized by analyzing $\langle T_{r}\rangle$ in the limit $r\to 0$ and imposing the condition $\langle T_{\delta r} (u)\rangle<\langle T_0(u)\rangle$. This yields the well-known criterion $CV_0(u)>1$, where $CV_0(u)=\sigma [T_0(u)]/\langle T_0(u)\rangle$  is the coefficient of variation of the unconditional FPT density associated with the underlying resetting-free dynamics. In terms of the first and second moments of the unconditional FPT, this condition can be recast as the inequality
\begin{equation}\label{cv0}
    \mathcal{H}(n,\Pi_0,u,l)>0.
\end{equation}
The explicit expression of $\mathcal{H}$ is deferred to Appendix~\ref{cvss}. To characterize the effect of target reactivity, we introduce the reactivity ratio $n=\Pi_l/\Pi_0$ and plot $\mathcal{H}(n,\Pi_0,u,l)$ in the $(u,n)$ parameter space as shown in Fig.~\ref{fig2}(a). The white contours, referred to as critical lines and determined by $\mathcal{H}(n,\Pi_0,u,l)=0$, delineate the transition between two regimes: one in which stochastic resetting enhances the performance of the search process, $R_{\rm E}$, and another in which resetting becomes detrimental, $R_{\rm D}$.

Let us first consider the parameter regime $R_{\rm D}$ enclosed by the critical lines. In this regime, stochastic resetting returns the particle to its initial position, which is typically farther away from the absorbing boundaries, thereby hindering the completion of the search process. By contrast, in the regime $R_{\rm E}$, trajectories that diffuse away from the targets towards the central region are reset back to their initial position located closer to the boundaries, which enhances the efficiency of the search process. The behavior at low values of the reactivity ratio $n$ can be understood by the asymmetric nature of the target reactivities. Small values of $n$ correspond to a highly reflective target at $x=l$ and a highly absorbing target at $x=0$, implying the completion of the search process is predominantly governed by absorption at $x=0$. Consequently, trajectories originating near the boundary at $x=l$ are more likely to terminate only after reaching the absorbing target at $x=0$. In this parameter regime, resetting repeatedly returns the particle away from the absorbing boundary, thereby impeding the completion of the search process.

For a fixed value of the initial position $u$, the critical lines determine a critical reactivity ratio $n_c$. Setting $u=0.4$, the corresponding unconditional MFPTs (Eq.~\eqref{mfpt-uncond-res-main}) are plotted in Figure~\ref{fig2}(b) as functions of resetting rate $r$ for different values of $n$. As the reactivity ratio $n$ decreases, the unconditional MFPT develops a global minimum at a finite optimal resetting rate $r_*$. Fig.~\ref{fig2}(c) shows that $r_*$ decreases to zero beyond a critical reactivity ratio $n_c$ [also marked in Fig.~\ref{fig2}(a) corresponding to $u=0.4$], marking a transition in the efficiency of stochastic resetting.
While the inequality in Eq.~\eqref{cv0} imposes a strong constraint on the system parameters, including the initial condition, it does not fully capture the impact of resetting on conditional FPTs.

\subsection{Conditional first-passage process in presence of stochastic resetting}
  \begin{figure*}
    \centering
    \includegraphics[width=\linewidth]{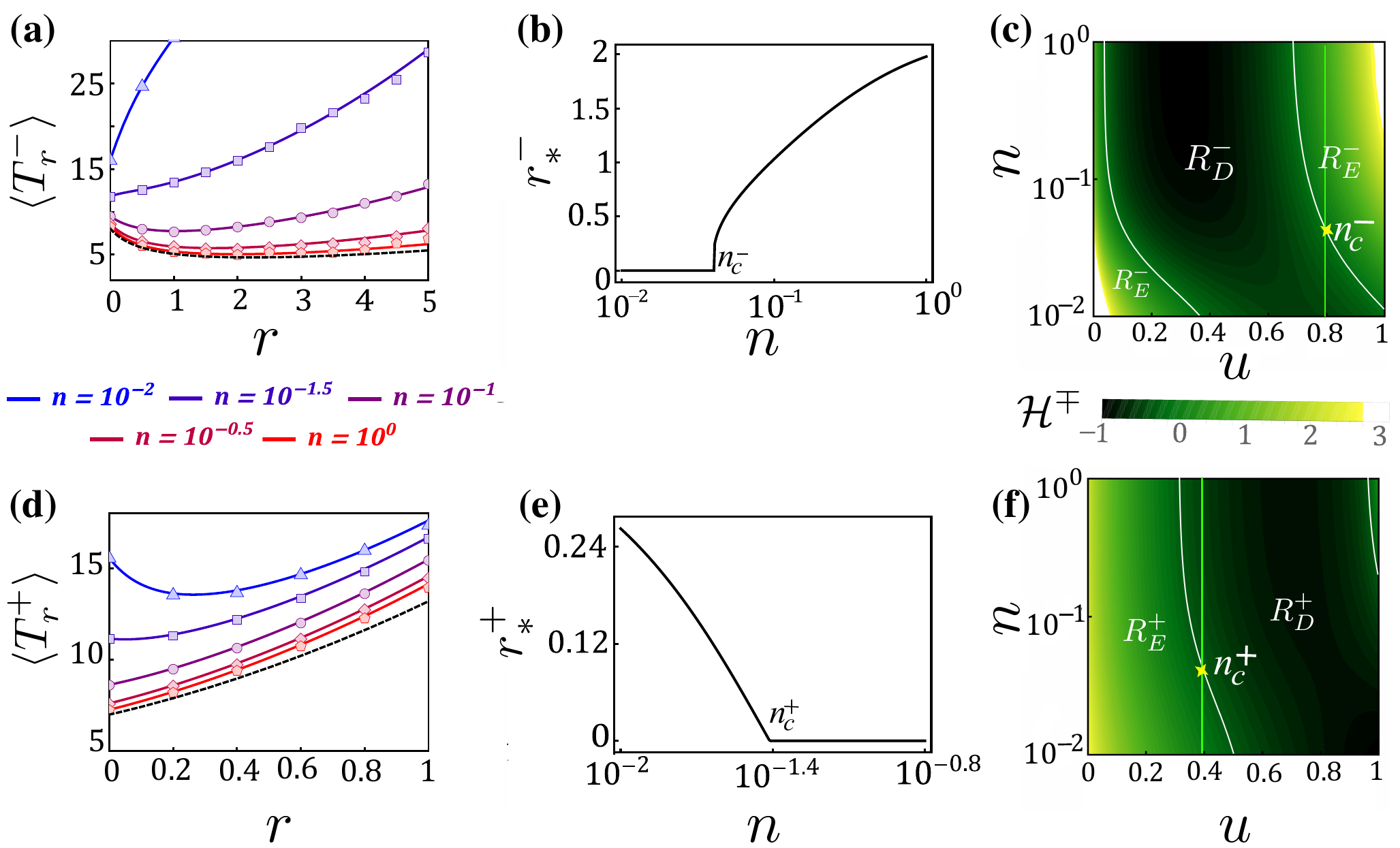}
    \caption{
    Conditional first-passage processes for exit through boundaries at $x=0$ (upper panels) and $x=l$ (lower panels). 
    (a, d) Conditional mean first passage times (MFPTs) as functions of the resetting rate for different values of reactivity ratio $n=\Pi_l/\Pi_0$.
    (b, e) Optimal resetting rates as functions of the reactivity ratio. 
    (c, f) Heat maps of $\mathcal{H}^\mp(n,\Pi_0,u,l)$. The white contours correspond to $\mathcal{H}^\mp(n,\Pi_0,u,l)=0$, or equivalently $CV^\mp_0(u)=\Lambda^\mp_0(u)$. These contours separate the parameter space into the resetting-enhanced regime $R_{\rm E}^{\mp}$ and the detrimental-resetting region $R_{\rm D}^{\mp}$. 
    In (c, f), the initial positions are chosen as $u=0.8$ and $u=0.4$, respectively. The corresponding conditional MFPTs, $\langle T^{\mp}_r\rangle$, are shown in panels (a) and (d), while the associated optimal resetting rates, $r^{\mp}_*$, are displayed in panels (b) and (e). Parameters: $l=5,~D=0.5$, and $\Pi_0=10$.}
    \label{fig3}
\end{figure*}

We begin by introducing the probability fluxes $J^\mp_r(u,t)$ associated with left (-) and right (+) boundaries. Owing to stochastic resetting at rate $r$, these fluxes obey a renewal structure analogous to Eq.~\eqref{surv-renew}, leading to~\cite{pal2019first,pal2025universal}
\begin{align} \label{renew-flux}
    J_r^\mp(u,t)= &e^{-rt}J_0^\mp(u,t)\nonumber\\
    &+r\int_0^t d\tau~e^{-r\tau}~Q_r(u,t-\tau)~J_0^\mp(u,\tau), 
\end{align}
where $J_0^\mp(u,t)$ is the probability flux through the corresponding boundary  for the underlying resetting-free process. The renewal structure separates the probability current into two distinct contributions: the first term in  Eq.~\eqref{renew-flux} corresponds to the trajectories that terminate without any resetting events up to time $t$, while the second term accounts for all trajectories involving one or more resetting events prior to termination. Using Eq.~\eqref{surv-renew-lap}, the probability fluxes in Eq.~\eqref{renew-flux} are obtained in the Laplace domain,
\begin{equation}\label{renew-flux-lap}
    \widetilde{J}_r^\mp(u,s)= \dfrac{\widetilde{J}_0^\mp(u,r+s)}{1-r\widetilde{Q}_0 (u,r+s)},
\end{equation}
and the corresponding escape probabilities through the boundaries can be expressed as
\begin{equation}\label{escape-res-1}
    \epsilon_r^\mp(u)= \int_0^\infty J_r^\mp(u,t)~dt=\widetilde{J}^{\mp}_r(u,s\to 0),
\end{equation}
which, using Eqs.~\eqref{fpt0-minus} and \eqref{fpt0-plus}, reduces to
\begin{equation}\label{escape-res-2}
    \epsilon_r^\mp(u) = \dfrac{\widetilde{J}_0^\mp(u,r)}{1-r\widetilde{Q}_0 (u,r)}= \epsilon_0^\mp(u)\dfrac{\widetilde{T}_0^\mp(u,r)}{\widetilde{T}_0(u,r)}.
\end{equation}
Recalling the relation, $f_{T_r^\mp}(u,t)= J_r^\mp (u,t)/\epsilon_r^\mp(u)$, the conditional FPT densities in the presence of resetting can be expressed in terms of the underlying Laplace-transformed  FPT densities (see Eqs.~\eqref{fpt0-minus}, \eqref{fpt0-plus} and \eqref{fpt0}) as
\begin{equation}\label{cond-den}
    \widetilde{T}^\mp_r(u,s)=\dfrac{\widetilde{T}^{\mp}_0(u,r+s)}{\widetilde{T}^\mp_0(u,r)}\dfrac{(s+r)\widetilde{T}_0(u,r)}{s+r\widetilde{T}_0(u,r+s)}.
\end{equation}
The mean conditional FPT then takes the form
\begin{align}
    \langle T^{\mp}_r(u)\rangle =& \langle T_r(u)\rangle+\dfrac{\partial}{\partial r}\ln\left[\dfrac{\widetilde{T}_0\left(u,r\right)}{\widetilde{T}^\mp_0\left(u,r\right)}\right].\label{cond-1st-moment}
\end{align}
Substituting Eqs.~\eqref{fpt0-minus},\eqref{fpt0-plus}, \eqref{fpt0} and \eqref{mfpt-uncond-res-main} in Eq.~\eqref{cond-1st-moment}, the conditional MFPT in the presence of resetting reduces to
\begin{widetext}
    \begin{align}
    \langle T_r^- (u)\rangle = &\langle T_r (u)\rangle+ \dfrac{(D+\Pi_0ul)}{(D(\Pi_0+\Pi_l)+\Pi_0\Pi_l l)}\partial_r\log\left[\dfrac{\mathcal{A}(u,\Pi_0,\Pi_l)}{\mathcal{B}(1-u,\Pi_l)}\right],\\
    \langle T_r^+ (u)\rangle = &\langle T_r (u)\rangle+ \dfrac{(D+\Pi_l(1-u)l)}{(D(\Pi_0+\Pi_l)+\Pi_0\Pi_l l)}\partial_r\log\left[\dfrac{\mathcal{A}(u,\Pi_0,\Pi_l)}{\mathcal{B}(u,\Pi_0)}\right],
\end{align}
with $\mathcal{A}(u,\Pi_0,\Pi_l) = (\alpha(r)D\Pi_0\cosh{[(1-u)l\alpha(r)]}+\alpha(r)D\Pi_l\cosh{[ul\alpha(r)]}+\Pi_0\Pi_l(\sinh{[ul\alpha(r)]}+\sinh{[(1-u)l\alpha(r)]}))$ and $\mathcal{B}(u,\Pi)=(\alpha(r)D\cosh{[ul\alpha(r)]}+\Pi\sinh{[ul\alpha(r)]})$.
\end{widetext}
In the limit $\Pi_0\to \infty$ and $\Pi_l\to \infty$, we approach~\cite{pal2019first,pal2025universal} 
\begin{align}
    \langle T_r^-(u)\rangle &= \langle T_r(u)\rangle + \dfrac{1}{r}\dfrac{e^{l\alpha_0}}{e^{2l\alpha_0}-e^{2ul\alpha_0}}\mathcal{F}(u,l),
    \label{uncond-mfptr-abs-plus}\\
    \langle T_r^+(u)\rangle &= \langle T_r(u)\rangle - \dfrac{1}{r}\dfrac{1}{e^{2ul\alpha_0}-1}\mathcal{F}(u,l)\label{uncond-mfptr-abs-min}
\end{align}
where
\begin{equation}
    F(u,l)=\dfrac{l\alpha_0\left[e^{2l\alpha_0}-e^{4ul\alpha_0}+(1-2u)e^{2ul\alpha_0}(1-e^{2l\alpha_0})\right]}{2(e^{l\alpha_0}-1)(e^{l\alpha_0}+e^{2ul\alpha_0})}\nonumber.
\end{equation}

Figures~\ref{fig3}(a) and \ref{fig3}(d) illustrate the dependence of the conditional MFPTs $(\langle T_r^{\mp}\rangle)$ on the resetting rate for different values of the reactivity ratio $n=\Pi_l/\Pi_0$. Depending on $n$, the conditional MFPTs display either monotonic or non-monotonic behavior with resetting, leading to either vanishing or finite optimal resetting rates ($r_*^{\mp}$). As the reactivity ratio is varied, the optimal resetting rate undergoes a transition from zero to a non-zero value, thereby defining a critical reactivity ratio ($n_c^{\mp}$) as shown in Figs.~\ref{fig3}(b) and \ref{fig3}(e). It is worth emphasizing here that the existence of finite optimal resetting rates $r_*^{\mp}$ can facilitate the search process. As illustrated in Figs.~\ref{fig3}(b) and \ref{fig3}(e), there exist regimes of reactivity ratio $n$ in which resetting decreases the conditional MFPT $\langle T_r^+\rangle$ associated with the target at $u=1$, while increasing $\langle T_r^-\rangle$ associated with the target at $u=0$ and vice versa. A deeper understanding of the efficiency of stochastic resetting for first passage processes to the left ($u=0$) and right ($u=1$), the targets can be examined by analyzing $\langle T_r^{\mp}\rangle$ in the limit $r\rightarrow 0$ and imposing the condition $\langle T_{\delta r}^{\mp}\rangle<\langle T_0^{\mp}\rangle$. This leads to a criterion distinct from that of the unconditional MFPT as,
\begin{equation}\label{def-cv0-lam0}
    CV^\mp_0(u)>\Lambda^\mp_0(u).
\end{equation}
Here $ CV^\mp_0(u)$ is the coefficient of variation for the conditional mean first passage time, defined as
\begin{align}
    CV_0^\mp(u)=\frac{\sqrt{\langle(T^\mp_0(u))^2\rangle-\langle T^\mp_0(u)\rangle^2}}{\langle T^\mp_0(u)\rangle},
\end{align}
and the threshold $\Lambda^\mp_0(u)$ is given by
\begin{align}
    \Lambda_0^\mp(u)=\frac{\langle T_0\rangle}{\langle T^\mp_0\rangle}\sqrt{\frac{1}{2}[1+CV^2_0(u)]}.
\end{align}

Substituting Eqs.~\eqref{mfpt0-minus}, \eqref{mfpt0-plus} and \eqref{mfpt0} into Eq.~\eqref{def-cv0-lam0} yields the following inequality:
\begin{equation}\label{cond-cv0}
    \mathcal{H}^\mp(n,\Pi_0,u,l)>0.
\end{equation}

For brevity, the explicit form of $\mathcal{H}^\mp(n,\Pi_0,u,l)$ is relegated to Appendix~\ref{cvss}. In Figs.~\ref{fig3}(c) and \ref{fig3}(f), we plot  $\mathcal{H}^\mp(n,\Pi_0,u,l)$ over the parameter space spanned by reactivity ratio $n$ and initial position $u$ for the left and right boundaries, respectively. 
The white curves in the heat maps denote the critical lines satisfying $\mathcal{H}^\mp(n,\Pi_0,u,l)=0$ or equivalently $CV^\mp_0(u)=\Lambda^\mp_0(u)$. Similar to the previous section, the critical lines divide the parameter space into two distinct regimes: one in which resetting enhances the efficiency of the search process, denoted by $R_{\rm E}^{\mp}$, and another in which resetting becomes detrimental, denoted by $R_{\rm D}^{\mp}$. We now discuss two heat maps separately.

We first consider the first passage process conditioned on exit through the boundary at $x=0$ [see Fig.~\ref{fig3}(c)]. For larger values of $n$, both boundaries become nearly equally absorbing, leading to a competition between the reactivity ratio $n$ and initial position $u$ in determining whether stochastic resetting expedites the completion of the search process. In contrast, for smaller values of $n$, the boundary at $x=l$ becomes highly reflective. Consequently, trajectories originating near this boundary can typically reach the absorbing boundary faster at $x=0$ without the aid of resetting. Introducing resetting for such trajectories repeatedly relocates the particle away from the absorbing boundary, thereby delaying the completion of the process. This explains why the detrimental-resetting region $R_{\rm D}^-$ is extended up to $u\approx 1$ at low $n$ values. On the other hand, trajectories initiated near the boundary at $x=0$ benefit from resetting  since resetting suppresses excursions  away from the absorbing boundary and thus accelerates the completion of the search process.

We next consider the first passage process conditioned on exit through the boundary at $x=l$ [see Fig.~\ref{fig3}(f)]. For all values of $n$, trajectories initiated within the region $R_{\rm E}^+$ benefit from resetting, as resetting curtails excursion towards the opposite boundary at $x=0$ and promotes exit through the boundary at $x=l$.
On the other hand we encounter a situation that cannot be fully explained by the criterion in Eq.~\eqref{cond-cv0}. 
For small values of $n$, the boundary at $x=l$ is highly reflective, such that absorption requires repeated interactions with the boundary over long time scales. Under the present resetting protocol, trajectories initiated near $x=l$ benefit from resetting, since repeated relocation to the initial position localizes the  particle near the boundary and thereby increases the probability of eventual absorption through the right boundary. However, according to the CV criterion [Eq.~\eqref{cond-cv0}], this region belongs to the detrimental-resetting regime, implying that resetting should not enhance the search process toward the boundary at $x=l$.
This observation emphasizes that the CV criterion [Eq.~\eqref{cond-cv0}] constitutes a necessary, but not sufficient, condition for resetting-enhanced search, consistent with earlier studies ~\cite{reuveni2016optimal,pal2017first}.  
In Figs.~\ref{fig3}(c) and \ref{fig3}(f), the initial positions are chosen as $u=0.8$ and $u=0.4$, respectively. The corresponding conditional MFPTs, $\langle T^{\mp}_r\rangle$,  shown in Figs.~\ref{fig3}(a) and \ref{fig3}(d), exhibit non-monotonic behavior as functions of the resetting rate, leading to finite optimal resetting rates for $n>n_c^-$ [Fig.~\ref{fig3}(b)] and  $n<n_c^+$ [Fig.~\ref{fig3}(e)]. 
\begin{figure}
    \centering
    \includegraphics[width=0.8\linewidth]{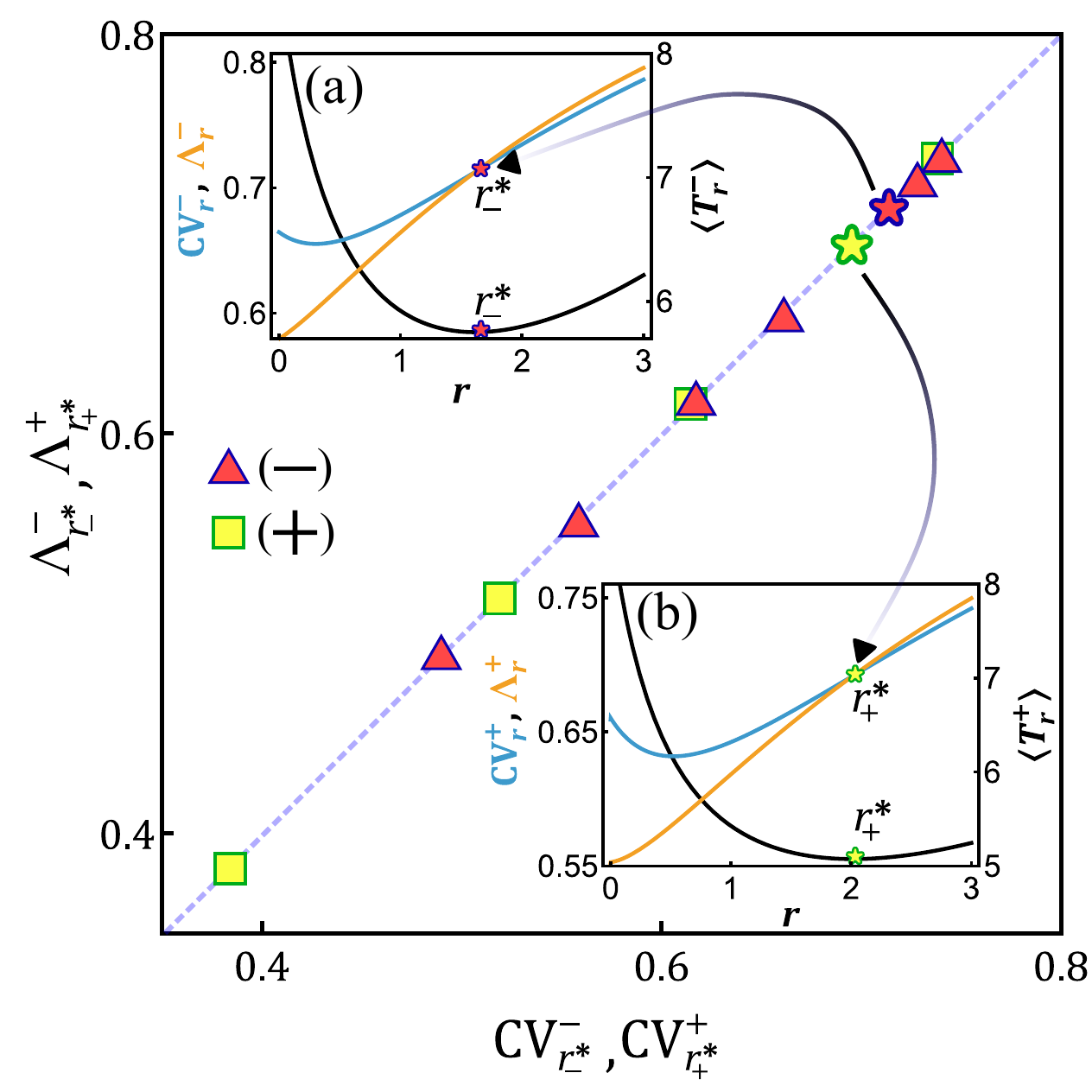}
    \caption{
    Verification of the relation for the conditional FPTs at the optimal resetting rate. The main panel shows $\Lambda^\mp_{r_\mp^*}(u)$ plotted against $CV^\mp_{r_\mp^*}(u)$, obtained by varying initial position $u$ while the system parameters are fixed at $n=10^{-0.5}$ with $\Pi_0=10$ and $l=5,~D=0.5$. The numerical data lie on a straight line with unit slope, thereby confirming the relation in Eq.~\eqref{cvr-lamr}. Insets illustrate the corresponding MFPTs together with $CV^\mp_{r}$ and $\Lambda^\mp_{r}$ as a function of the resetting rate. Inset (a) corresponds to the first passage process conditioned on exit through the left boundary with $u=0.8$, while inset (b) corresponds to exit through the right boundary with $u=0.2$. The markers denote the optimal resetting rates $r_\mp^*$, where the equality Eq.~\eqref{cvr-lamr} is satisfied. }
    \label{fig4}
\end{figure}

Let us now examine the relative fluctuations in conditional FPTs at the optimal resetting rate, which is defined as
\begin{align}
    CV_{r^*_{\mp}}^\mp(u)=\frac{\sqrt{\langle(T^\mp_{r^*_{\mp}}(u))^2\rangle-\langle T^\mp_{r^*_{\mp}}(u)\rangle^2}}{\langle T^\mp_{r^*_{\mp}}(u)\rangle}.
\end{align}
The second moment for the conditional FPT reads as 
\begin{align}\label{cond-2nd-moment}
    \langle (T^\mp_r(u))^2\rangle&=\langle T_r(u)^2\rangle\left[\widetilde{T}_0(u,r)+\dfrac{\langle T^{\mp}_r(u)\rangle}{\langle T_r(u)\rangle}(1-\widetilde{T}_0(u,r))\right]\nonumber\\
    &-\left[\dfrac{1}{\widetilde{T}_0(u,r)}\dfrac{\partial^2\widetilde{T}_0(u,r)}{\partial r^2}-\dfrac{1}{\widetilde{T}^{\mp}_0(u,r)}\dfrac{\partial^2\widetilde{T}^{\mp}_0(u,r)}{\partial r^2}\right].
\end{align}
The analytical expressions of $CV_{r}^\mp(u)$ are rather cumbersome and are therefore not presented explicitly. Instead, we plot $CV_{r}^\mp(u)$ as functions of the resetting rate in the insets (a) and (b) of Fig. \ref{fig4}. As evident from the figure, the coefficient of variation for the conditional FPTs satisfies
\begin{equation}\label{cvr-lamr}
    CV^\mp_{r_\mp^*}(u)=\Lambda^\mp_{r_\mp^*}(u),
\end{equation}
where the threshold $\Lambda^\mp_{r}$ is given by 
\begin{align}
    \Lambda_r^\mp(u)=\frac{\langle T_r(u)\rangle}{\langle T^\mp_r(u)\rangle}\sqrt{\frac{1}{2}[1+CV^2_r(u)]}.
\end{align}
While the proof of Eq.~\eqref{cvr-lamr} is given in \cite{pal2025universal}, its physical significance can be understood in terms of the reliability of the search process. Since $CV_r^{\mp}$ measures the relative fluctuations of the conditional FPTs, it quantifies the degree of reproducibility of the search process. This equality  Eq.~\eqref{cvr-lamr} therefore identifies the optimal resetting rate as the point where the reliability of the conditional search process attains a threshold set by the corresponding unconditional dynamics. Consequently, optimal resetting for conditional process is determined not only by the minimization of the conditional MFPT but also by a  balance between search efficiency and fluctuations. Thus optimal resetting strategy yields the fastest search toward a given target while preserving a specific degree of reliability.

\begin{figure}
    \centering
    \includegraphics[width=\linewidth]{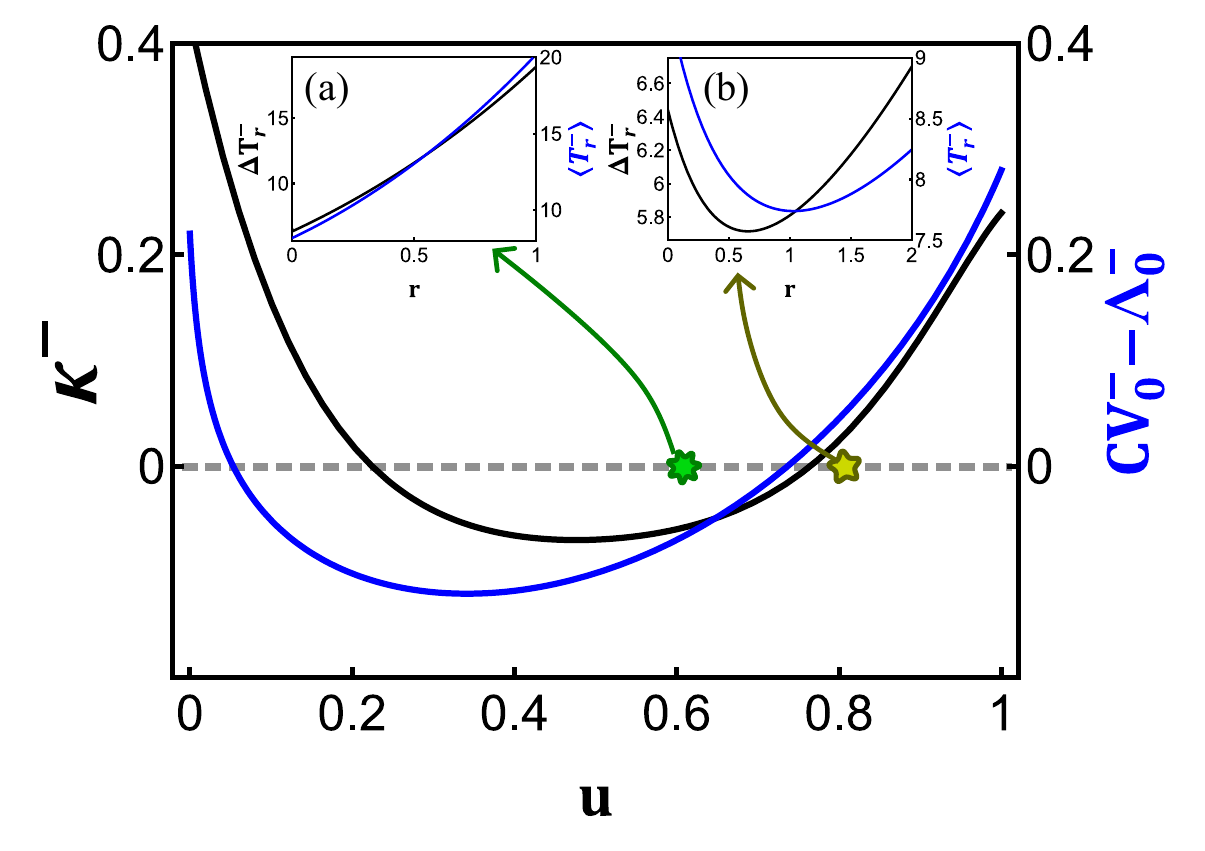}
    \caption{
    Parameter regimes characterizing the effectiveness of stochastic resetting in optimizing the conditional MFPT and its fluctuations for exit through the left boundary. The solid black and blue curves show $\kappa^-$ and $CV^-_0-\Lambda^-_0$, respectively, as functions of initial position $u$. We choose two representative initial position: $u=0.6$ (green marker), for which  $\kappa^-<0$ and $CV^-_0<\Lambda_0^-$, and $u=0.8$ (yellow marker), for which  $\kappa^->0$ and $CV^-_0>\Lambda_0^-$.The conditional MFPTs and their associated fluctuations as functions of resetting rate are shown in panels (a) and (b) for $u=0.6$ and $u=0.8$, respectively.
    }
    \label{kappa-minus}
\end{figure}

\section{Trade-off between conditional fluctuation and MFPTs}
Evidently, the efficiency of stochastic resetting in optimizing the conditional MFPTs can be assessed using the criterion in Eq.~\eqref{def-cv0-lam0} to optimize. Beyond the mean search time, however, the reliability of the resetting-mediated search process is encoded in the fluctuations of the conditional FPT densities. The fluctuations are characterized by variance
\begin{equation}\label{fluc-eq-main}
    (\Delta T^\mp_r)^2 = \langle (T^\mp_r(u))^2\rangle-\langle T^{\mp}_r(u)\rangle^2.
\end{equation}
Evaluating Eq.~\eqref{fluc-eq-main} in the limit $\delta r \to 0$, we obtain
\begin{align}\label{fluc-linear}
    (\Delta T^\mp_{\delta r\to 0})^2 &= (\Delta T^\mp_{0})^2-(\mu_2^\mp)^{3/2}\widetilde{\gamma}_1\kappa^{\mp}~\delta r\nonumber\\
    &~~~~~~~~~~~~~~~~~+\mathcal{O}((\delta r)^2),
\end{align}
where $\kappa^\mp =\gamma_1^\mp/\widetilde{\gamma}_1-\dfrac{1}{3}\left(\mu_2/\mu_2^\mp\right)^{3/2}$. Here, $\mu^\mp_2$ and $\mu^\mp_3$ denote the second and third central moments of the underlying FPT density $T^\mp_0(s)$, respectively. Furthermore, $\gamma^\mp_1$ represents the skewness of the conditional FPT distribution in the absence of resetting, whereas $\widetilde{\gamma}_1$ represents the standardized third raw moment of the corresponding unconditional FPT distribution $\widetilde{T}_0(s)$. Imposing the condition that an infinitesimal resetting rate reduces the fluctuations, \textit{i.e.,} $(\Delta T^\mp_{\delta r\to 0})^2<(\Delta T^\mp_{0})^2$, we obtain
\begin{equation}\label{kappa}
    \kappa^\mp>0.
\end{equation}

 Equation~\eqref{kappa} allows us to identify the parameter regimes in which stochastic resetting effectively reduces the associated fluctuations in conditional FPTs. For $u=0.6$, where $\kappa^-<0$ and $CV^-_0<\Lambda_0^-$, both the left--conditional MFPT, $\langle T_r^-\rangle$, and its fluctuation, $\triangle T_r^-$, increase monotonically with the resetting rate [see Fig.~\ref{kappa-minus} (a)]. Thus, resetting neither optimizes the conditional MFPT and nor reduces its fluctuation in this regime. In contrast, for $u=0.8$, where $\kappa^->0$ and $CV^-_0>\Lambda_0^-$, the conditional MFPT exhibits a non-monotonic dependence on the resetting rate, yielding a finite optimal resetting rate, while the associated fluctuations are simultaneously reduced [notice Fig.~\ref{kappa-minus}(b)]. These results demonstrate how the criteria in Eqs.~\eqref{def-cv0-lam0} and \eqref{kappa} distinguish the parameter regime in which stochastic resetting can optimize conditional search time and control its fluctuations. An analogous analysis for the resetting-mediated first passage process conditioned on exit through the right boundary is presented in the Appendix.~\ref{sec-kappa-plus}.

\section{Conclusions}
In this paper, we have undertaken a comprehensive investigation of the statistical properties of the first-passage process in a confined system with reactive boundaries. Although interfaces in physical, chemical, and biological systems are frequently modeled as either perfectly reflecting or perfectly absorbing, realistic boundaries generally exhibit partial reactivity due to microscopic processes such as adsorption and catalysis. To capture these effects, we studied the stochastic search process in a minimal model of a Brownian particle diffusing in one dimension between two partially reactive targets in the presence of instantaneous stochastic resetting. This setting provides a natural framework for elucidating the interplay between target reactivity and stochastic resetting in the search process, which is quantified and analyzed through two complementary measures, namely the unconditional MFPT and the conditional MFPT associated with each target. Our analysis demonstrates that the first-passage observables depend sensitively on the ratio of the reactivities of two targets. In particular, both the unconditional and conditional MFPTs exhibit either monotonic or non-monotonic dependence on the resetting rate and the ratio. Consequently, the optimal resetting strategy undergoes a 
transition between regimes of finite and vanishing optimal resetting rates. These findings reveal that the efficiency of stochastic resetting is not universal but is instead governed by the competition between targets of unequal reactivities. Finally, we examined the fluctuations of the conditional first-passage process and demonstrated that the optimal resetting condition is characterized by a universal criterion.

It is worth noting that our model provides a minimal framework for studying target search processes in chemical and biological systems where targets generally possess finite reactivity. While the first passage problems are often formulated in terms of idealized perfectly absorbing targets, realistic search processes can occur in complex environments where an encounter with a target does not necessarily lead to immediate absorption or reaction. Incorporating finite target reactivity is therefore crucial for capturing the kinetics for such imperfect search processes \cite{go2024active}. Our study provides a natural starting point in the field of resetting-mediated first passage dynamics with reactive targets and opens several avenues for future investigation, including extensions to active searchers, higher dimensions and complex geometries, heterogeneous target reactivities, and more realistic resetting protocols.


\section{Acknowledgment}
S.G. and S.P. equally contributed to this work. P.S.P.\ acknowledges research support from the Korea Institute for
Advanced Study through individual KIAS Grant No.~CG085601.

\bibliography{main}

\newpage
\onecolumngrid
\appendix

\section{Mathematical expressions of Eqs.~\eqref{cv0} and \eqref{cond-cv0} of the main text}\label{cvss}
In this section, we provide the explicit analytical expressions of $\mathcal{H}$ and $\mathcal{H}^{\pm}$. We begin with $\mathcal{H}$, which follows from  the unconditional coefficient of variation $CV_0$ in absence of resetting. The mean unconditional FPT , $\langle T_0(u)\rangle$, is given in Eq.~\eqref{mfpt0} while its second moment is obtained from $\langle (T_0(u))^2\rangle =\left.\partial_s^2\widetilde{T}_0(u,s)\right|_{s\to 0}$, with $\widetilde{T}_0(u,s)$ is given in Eq.~\eqref{fpt0}. Using $CV_0=\sqrt{\langle (T_0(u))^2\rangle-\langle T_0(u)\rangle^2}/\langle T_0(u)\rangle$ and combining the above expressions, we obtain
\begin{align}\label{uncond-cv0}
    \mathcal{H}(n,\Pi_0,u,l)&=\Pi_0l(\Pi_0^3l^3n^2u(1-6u+10u^2-5u^3)+4D^3(2-6u+3u^2+n(3u^2-1))\nonumber\\
    &+D\Pi_0^2l^2n(u(5-24u+30u^2-10u^3)+n(1-7u+6u^2+10u^3-10u^4))\nonumber\\
    &-D^2\Pi_0l(u(-8+24u-20u^2+5u^3)+n^2(1-6u^2+5u^4)\nonumber\\
    &+n(-5+28u-18u^2-20u^3+10u^4)))>0.
\end{align}
This essentially underpins the domain of system parameters, where resetting becomes profound in optimizing the unconditional MFPT as indicated in Fig.~\ref{fig2}.

Similarly, from the underlying conditional FPT densities to the left boundary, we find the first and second moments of the conditional FPTs. Further, using the inequality in Eq.~\eqref{def-cv0-lam0}, we find the domain of the system parameters in which resetting becomes pronounced in optimizing the conditional MFPT to the left boundary. Importantly, this domain can be manifested with the following inequality
\begin{align}
\mathcal{H}^-(n,\Pi_0,u,l)
&= \Pi_0 l \Big(
  60 D^5 (2-n-6u+3(1+n)u^2) \nonumber\\
&\quad
  + 15 D^4 \Pi_0 l \Big[
      (11-13n)n
      + 4(-2+n)(-1+6n)u
      + 6(-4+n(7+n))u^2 \nonumber\\
&\quad
    - 4(1+n)(-5+2n)u^3
      - 5(1+n)^2 u^4
  \Big] \nonumber\\
&\quad
  + 3 D^3 \Pi_0^2 l^2 n \Big[
      (17-38n)n
      + 5(11+n(-43+18n))u
      + 10(-24+n(35+2n))u^2 \nonumber\\
&\quad
      - 10(-31+5n(1+2n))u^3
      + 10(-14+n)(1+n)u^4
      + 18(1+n)^2 u^5
  \Big] \nonumber\\
&\quad
  + D^2 \Pi_0^3 l^3 n^2 \Big[
      (7-19n)n
      + 51u
      + 54(-5+n)n\,u
      + 15(-30+n(43+n))u^2 \nonumber\\
&\quad
      - 10(-95+2n(13+5n))u^3
      + 15(-50+(-24+n)n)u^4
      + 6(1+n)(37+9n)u^5
      - 19(1+n)^2 u^6
  \Big] \nonumber\\
&\quad
  - D \Pi_0^4 l^4 n^3 (-1+u) \Big[
      n+(7-44n)u+(-149+85n)u^2
      +2(163+10n)u^3 \nonumber\\
&\quad-2(107+50n)u^4
      +38(1+n)u^5
  \Big] - \Pi_0^5 l^5 n^4 (-1+u)^2 u
    \bigl(-1+u(24+u(-46+19u))\bigr)
\Big)>0. 
\end{align}
Furthermore, a similar analysis can be done for the resetting-mediated conditional first-passage through the right boundary. Corresponding sharp boundary can be expressed in terms of the following inequality
\begin{align}
\mathcal{H}^+(n,\Pi_0,u,l)
&= \Pi_0 l \Big(
  60 D^5 (2-n-6u+3(1+n)u^2) \nonumber\\
&\quad
  + \Pi_0^5 l^5 n^2 u^2
    \bigl(4-15u+30u^3-19u^4\bigr) \nonumber\\
&\quad
  - 15 D^4 \Pi_0 l \Big[
      -4 + (-3+n)n - 8u + 28nu
      - 6(-8+n(3+n))u^2 
      - 28(1+n)u^3
      + 5(1+n)^2 u^4
  \Big] \nonumber\\
&\quad
  - 3 D^3 \Pi_0^2 l^2 \Big[
      -n(8+3n)
      + 5(1+n)(-8+7n)u 
      + 10(4+(22-3n)n)u^2
      - 10(-12+7n(3+n))u^3 \nonumber\\
&\quad
      + 50(-2+n)(1+n)u^4
      + 18(1+n)^2 u^5
  \Big] \nonumber
  + D^2 \Pi_0^3 l^3 \Big[
      4n^2
      + 3n(16+n)u
      - 15(1+n)(-4+11n)u^2 \nonumber\\
&\quad
      + 30(-4+n(-10+7n))u^3
      + 75n(6+n)u^4 
      - 12(1+n)(-5+9n)u^5
      - 19(1+n)^2 u^6
  \Big] \nonumber\\
&\quad
  - D \Pi_0^4 l^4 n u \Big[
      -8n
      + 3(-8+7n)u
      + 75(1+n)u^2 
      - 150n u^3
      + 6(-15+4n)u^4
      + 38(1+n)u^5
  \Big]
\Big)>0.
\end{align}

\newpage
\renewcommand{\theequation}{B\arabic{equation}}
\renewcommand{\thefigure}{B\arabic{figure}}
\setcounter{equation}{0}
\setcounter{figure}{0}

\section{Fluctuations in resetting-mediated conditional FPTs through the right boundary}\label{sec-kappa-plus}
\begin{figure}[h]
    \centering
    \includegraphics[width=0.5\linewidth]{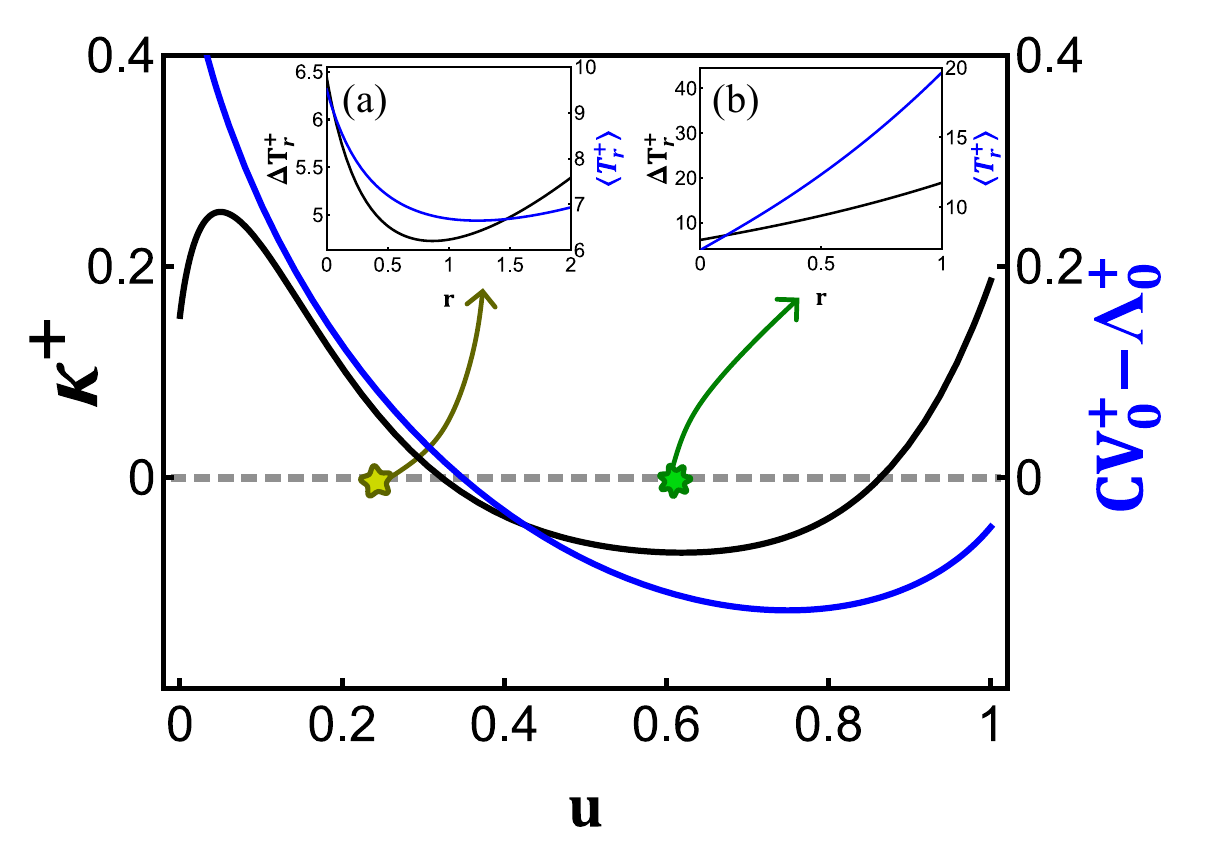}
    \caption{
    Parameter regimes characterizing the effectiveness of stochastic resetting in optimizing conditional MFPT and its fluctuations for exit through the right boundary. The solid black and blue curves show $\kappa^+$ and $CV^+_0-\Lambda^+_0$, respectively [see Eqs.~\eqref{kappa} and \eqref{def-cv0-lam0}], as functions of initial position $u$. Two representative initial positions are considered. For $u=0.24$ (yellow marker), $\kappa^+>0$ and the conditional MFPT, $\langle T^+_r(u)\rangle$, exhibits a non-monotonic dependence on the resetting rate, yielding a finite optimal resetting rate while the associated fluctuation, $\Delta T^+_r(u)$, is simultaneously reduced, as shown in panel (a). For $u=0.6$ (green marker), both $\langle T^+_r(u)\rangle$ and $\Delta T^+_r(u)$ increase with the resetting rate, as shown in panel (b). These results illustrate the criteria in Eqs.~\eqref{def-cv0-lam0} and \eqref{kappa} for identifying the parameter regimes in which stochastic resetting optimizes the conditional MFPT and reduces its fluctuations.
    }
    \label{kappa-plus}
\end{figure}

Here we discuss the effect of resetting on the trade-off between fluctuations and mean value of the conditional FPTs through the right boundary. For $u=0.24$, where $\kappa^+>0$ and $CV^+_0>\Lambda_0^+$, the conditional MFPT exhibits a non-monotonic dependence on the resetting rate, yielding a finite optimal resetting rate, while the associated fluctuations are simultaneously reduced [notice Fig.~\ref{kappa-plus}(a)]. In contrast, for $u=0.6$, where $\kappa^+<0$ and $CV^+_0<\Lambda_0^+$, both the right--conditional MFPT, $\langle T_r^+\rangle$, and the corresponding fluctuation, $\Delta T_r^+$, increase monotonically with the resetting rate [see Fig.~\ref{kappa-plus} (b)]. Thus, resetting neither optimizes the conditional MFPT and nor reduces its fluctuation in this regime.  

\end{document}